# Crowdsourcing with Endogenous Entry

Arpita Ghosh[*]     Preston McAfee[†]

September 7, 2018

**Abstract**

We investigate the design of mechanisms to incentivize high quality outcomes in crowdsourcing environments with strategic agents, when entry is an *endogenous*, strategic choice. Modeling endogenous entry in crowdsourcing markets is important because there is a nonzero cost to making a contribution of any quality which can be avoided by not participating, and indeed many sites based on crowdsourced content do not have adequate participation. We use a mechanism with monotone, rank-based, rewards in a model where agents strategically make participation and quality choices to capture a wide variety of crowdsourcing environments, ranging from conventional crowdsourcing contests with monetary rewards such as TopCoder, to crowdsourced content as in online Q&A forums.

We begin by explicitly constructing the unique mixed-strategy equilibrium for such monotone rank-order mechanisms, and use the participation probability and distribution of qualities from this construction to address the question of designing incentives for two kinds of rewards that arise in the context of crowdsourcing. We first show that for attention rewards that arise in the crowdsourced *content* setting, the entire equilibrium distribution and therefore every increasing statistic including the maximum and average quality (accounting for participation), improves when the rewards for every rank but the last are as high as possible. In particular, when the cost of producing the lowest possible quality content is low, the optimal mechanism displays all but the poorest contribution. We next investigate how to allocate rewards in settings where there is a fixed total reward that can be arbitrarily distributed amongst participants, as in crowdsourcing contests. Unlike models with exogenous entry, here the expected number of participants can be increased by subsidizing entry, which could potentially improve the expected value of the best contribution. However, we show that subsidizing entry does not improve the expected quality of the best contribution, although it may improve the expected quality of the average contribution. In fact, we show that free entry is dominated by taxing entry— making all entrants pay a small fee, which is rebated to the winner along with whatever rewards were already assigned, can improve the quality of the best contribution over a winner-take-all contest with no taxes.

[*]Yahoo! Research, Santa Clara, CA. arpita@yahoo-inc.com
[†]Yahoo! Research, Burbank, CA. mcafee@yahoo-inc.com

# 1 Introduction

Crowdsourcing, where a problem or task is broadcast to a crowd of potential contributors for solution, is a rapidly growing online phenomenon being used in applications ranging from seeking solutions to challenging projects such as in Innocentive or TopCoder, all the way to crowdsourced *content* such as on online Q&A forums like Y! Answers, StackOverflow or Quora. The two key issues which arise in the context of crowdsourcing are quality— is the obtained solution or set of contributions of high quality?— as well as *participation*— there is a nonzero effort or cost associated with making a contribution of *any* quality in a crowdsourcing environment which can be avoided by simply choosing to not participate, and indeed many sites have too little content. In such a setting, the effort an agent decides to exert will depend on how many other agents are likely to participate and how much effort they will exert, since the amount of effort necessary to obtain a particular reward depends both on the number and strength of competitors an agent faces. Naturally, the level of effort agents choose, and therefore the quality of the output created, depends on the incentives offered to agents. How should rewards be designed to incentivize high effort, in a setting where entry is an *endogenous*, strategic choice?

We are motivated by two different kinds of questions that arise in the context of designing rewards for crowdsourced content, depending on the setting and the nature of the rewards. The first is in the context of attention rewards on user-generated content (UGC) based sites, such as online Q&A forums like Quora or StackOverflow. Here, the mechanism designer, or site owner, has a choice about how many of the received contributions to display, *i.e.*, how to reward the contributions with attention — he could choose to display all contributions for a particular task, or display only the best few, suppressing some of the poorer contributions. What strategy improves the quality of the best contribution supplied? What about the average quality of contributions? On the one hand, suppression should cause quality to rise, because the payoff to poor content falls; on the other hand, suppressing content also corresponds to decreasing the total reward paid out, which could decrease quality. Is it a good idea, in a game-theoretic sense, to display all contributions?

The second question arises in settings where there is some fixed total available reward which can be distributed arbitrarily amongst the agents. This happens, for example, in the setting of crowdsourcing contests with monetary rewards, where the principal posing the challenge or task offers some fixed amount of money for obtaining the solution to the challenge. Another instance is systems which reward agents with virtual points[1]. (The distinction between

---

[1]If the value of points is determined only in proportion to the total number of points awarded (so that just doubling the number of points awarded for all tasks has no effect on

this setting and attention rewards is that it is not possible to take away attention from the second position and add it to the first position since, to a first approximation, attention to the second spot comes from a subset of viewers providing attention to the first; so attention rewards cannot be arbitrarily redistributed across ranks.) How can rewards be designed to improve the quality of contributions in settings with arbitrarily redistributable rewards, when entry is endogenous?

**Our contributions.** We use a mechanism with monotone, rank-based rewards in a model with contributors who strategically choose both participation and quality to simultaneously capture a wide variety of crowdsourcing environments, ranging from conventional crowdsourcing contests with monetary rewards such as TopCoder, to crowdsourced content such as in Q&A forums. We first analyze the equilibria of such monotone rank-order mechanisms, and explicitly construct the unique mixed-strategy equilibrium for this mechanism (§3). We then use this construction, which explicitly gives us the equilibrium participation probability and distribution of qualities, to address the question of how to design rewards for each of the two settings previously mentioned.

We first show (§4) that for attention rewards, the entire equilibrium distribution and therefore every increasing statistic, including the maximum and average quality (accounting for participation) improves when the rewards for every rank but the last are as high as possible: if there are $n$ potential contributors, then the optimal mechanism sets the attention rewards for ranks 1 through $n-1$ to be the maximum possible, while the attention to the $n$th rank is curtailed to the cost of producing the lowest possible quality contribution (note here that $k < n$ agents may participate, in which case only the rewards for ranks $1, \ldots, k$ are given out). If this cost is low, this prescribes, roughly speaking, displaying all but the poorest contribution.

We next investigate redistribution of rewards (§5). Unlike in models with exogenous entry with a fixed number of participants, it is possible here to increase the expected number of participants by subsidizing entry, for example, by providing a small reward to all participants in addition to a large reward to the winner. In models with exogenous entry, more participants lead to higher qualities, suggesting that subsidizing entry may be productive in this endogenous entry setting as well. Also, even if subsidizing entry (at the cost of paying less to the winner) were to reduce the equilibrium distribution from which each contributor chooses her quality, the *expected* value of the maximum quality could nonetheless increase when the number of contributors increases, since we have the maximum of a larger number of random variables. However, we show that subsidizing entry does not improve the expected value of the maximum quality, although it may improve the expected value of the total contribution. In fact, we show that free entry (corresponding to a winner-take-all contest) is dominated by *taxing* entry - making all entrants pay a small fee, which is rebated

---

incentives), the total number of points available to reward agents with is effectively fixed as well.

to the winner along with whatever rewards were already assigned, can improve the expected quality of the outcome.

**Related work.** There is a growing literature on the optimal design of contests [13, 14, 12], as well as specifically on the design of online crowdsourcing contests [1, 4, 3] and online procurement (e.g., [15]). The most relevant of these to our work are the following. [13] investigates the optimal structure of rewards when the objective is to maximize the sum of qualities of contributions, for concave, linear and convex costs; [14] also considers the objective of maximizing the highest quality contribution. [1] studies the optimal design of crowdsourcing contests in a setting with agents with heterogeneous abilities and linear costs, when the objective is to maximize the sum of the top $k$ qualities minus the total reward paid out to agents. [3] study the design and approximation of optimal crowdsourcing contests modeled as all-pay auctions, again for agents with linear costs, and investigate the extent of wasted effort compared to conventional procurement. There is also a voluminous economics literature on contest design not focused on crowd-sourcing, see, for example, [2] and references therein.

The key difference between this literature and our work is *endogenous entry*– all these papers assume some fixed number $n$ of contestants who always participate (*i.e.*, the cost of producing the lowest possible quality $c(0) = 0$), whereas whether to participate or not is an endogenous strategic choice in our model (*i.e.*, we allow for $c(0) > 0$). That endogenous participation may matter is foreshadowed by the auction literature, which is the basis for much of the modeling of crowdsourced content provision— auctions with endogenous entry are quite different than auctions with exogenous participation. For instance, while posting monopoly reserve prices is always part of seller maximization in auction models with exogenous participation, a monopoly seller sets efficient reserve prices when participation is endogenous [11]. Endogenous entry makes a substantial difference in the crowd-sourcing models for much the same reason— it is no longer possible to reduce the profits of the contributors, because those profit levels are determined by the cost of entry. Our results on the optimality of taxing in §5 are foreshadowed by Taylor [16] and Fullerton and McAfee [5], both of whom show, albeit in different settings, that free entry produces too much entry. An additional, though less important, difference with the literature on online crowdsourcing contests [1, 3] is that we allow general cost functions rather than restricting linear cost functions.

There is also a related literature on models with endogenous entry [11, 5, 16, 9, 7, 8], although largely outside the specific setting of contest design (with the exception of [16, 8]). [16] studies a setting with agents who all have a common exogenous cost to participation, and draw the quality of their output from some distribution. An agent's only strategic choice is whether or not to enter in each period of a possibly multi-period game. In this model, [16] finds that restricting entry with taxation is optimal. The key difference from our work, of course, is that quality is an endogenous choice in our model as opposed to an exogenous draw from a distribution. [8] uses a very similar agent model to that in [16], but instead addresses the question of implementation of optimal outcomes— are contest structures where the highest quality contribution receives some high prize and all other contributions receive some low prize adequate to implement the optimal outcome achievable with nonstrategic agents? We do not address the question of implementability of optimal outcomes, but rather ask how to improve equilibrium outcomes. [9, 7] address the question of incentivizing high-quality user-generated content (UGC) in a game-theoretic framework with strategic agents and endogenous entry, a setting related to that of crowdsourced content. However, [9, 7] focus on the performance of mechanisms in the limit of diverging rewards (as is the case with attention rewards in the context of very popular UGC sites such as Youtube or Slashdot), while our results address the case of finite, or bounded, rewards, as is relevant in much of crowdsourcing.

## 2 Model

We model a general social computing or crowdsourcing scenario as a game with rank-dependent rewards, *i.e.*, a rank-order mechanism with reward $a_i$ for producing the $i$th best contribution, and focus on the effect of the reward structure on the qualities of contributions produced by strategic agents in a single micromarket, such as one crowdsourcing contest or a question in a Q&A forum.

There is a micromarket with a pool of $n$ agents, each of whom is a potential participant in this micromarket. Each agent can choose whether to contribute or not, as well as the quality of the contribution she makes if she chooses to enter. Agents make the decision of whether to participate strategically, *i.e.*, entry is *endogenous*, and each agent that chooses to participate then chooses her level of effort, modeled as the quality $q$ of the output she produces, strategically to maximize her utility. We next describe the utility of an agent.

The cost, or effort, required to producing a contribution of quality $q$ for each agent is $c(q)$. We will assume that $c(q) \geq 0$ is some strictly increasing, continuously differentiable function of $q$. Although we do not need this assumption, it will be useful to imagine that $c(0) > 0$, *i.e.*, there is a nonzero cost to producing a contribution, even one of the lowest possible quality. This nonzero participation cost models, for example, the cost of reading and understanding the task for which contributions are being solicited, which can be avoided by simply choosing to not participate. Since $c(0) > 0$, participation always has a strictly positive cost, whereas not participating at all incurs zero cost and produces zero benefit, and therefore has a net utility of 0.

*Homogeneity.* Note that our model of costs assumes *homogeneity* amongst all potential contributors, corresponding to assuming that agents do not differ in their abilities, but simply in the amount of effort that they choose to put in. While there are indeed settings where potential contributors may differ widely in their abilities, there are also settings where it is effort, rather than ability, which dominates the quality of the outcome produced (for example, writing a review for a product on Amazon, or producing an article for a crowdsourced-content based site such as Associated Content which requires exhaustively researching the topic rather than inherent expertise, fall in this category). Also, in several settings, such as specific topics or categories in Q&A forums like Quora or Stackoverflow, the set of potential contributors may be self-selected to have rather similar abilities or expertise levels, and therefore have similar costs to producing a particular quality. While the most complete model of the real world would allow for differences in both ability and effort, we choose here to focus on strategic choice of effort, *i.e.*, to focus on the strategic question faced by an agent of "how little effort can I get away with?", since this is a reasonable first approximation in many settings relevant to crowdsourcing.

**Mechanism $\mathcal{G}(a_1, a_2, \ldots, a_n)$.** Once agents have made their participation and quality choices, the mechanism observes the qualities $q_i$ produced by the agents who enter, and awards prizes $a_i$ to the participants in decreasing order of quality. Specifically, a mechanism $\mathcal{G}(a_1, a_2, \ldots, a_n)$ awards a prize of value $a_i$ to the entrant who produces the $i$th highest-quality contribution. If more than one agent produces the same quality, the mechanism breaks ties randomly amongst these agents to obtain a strict rank order, and assigns rewards according to this order. No prizes are awarded to agents who do not enter, and specifically, if no agent participates, no prize is awarded. We note here that we assume that qualities are perfectly observable, as in all the prior literature on contest design and crowdsourcing contests [13, 3, 4, 1]— since each task in a crowdsourcing environment is usually posed by some principal who can rank the contributions in decreasing order of quality (such as the person posting the task in a crowdsourcing contest or the asker in an online Q&A forum), this assumption is reasonable, particularly since $\mathcal{G}(a_1, a_2, \ldots, a_n)$ only uses the relative ranks, and not the actual absolute values of the qualities.

We will focus throughout on *monotone* mechanisms, in which higher ranks receive higher rewards, and not all rewards are equal.

**Definition 2.1.** *Consider a micromarket with n agents. We say $\mathcal{G}(a_1, a_2, \ldots, a_n)$ is a monotone mechanism if $a_1 \geq a_2 \ldots \geq a_n$ and at least one inequality is strict, i.e., $a_i > a_{i+1}$ for some $1 \leq i \leq n-1$. We say $\mathcal{G}(a_1, a_2, \ldots, a_n)$ is monotone nonnegative if $\mathcal{G}(a_1, a_2, \ldots, a_n)$ is monotone and $a_n \geq 0$, i.e., all rewards are nonnegative.*

**Solution concept.** We use the solution concept of a symmetric Nash equilibrium, since agents' payoff functions are symmetric in the parameters of the game. In a symmetric strategy, each contributor participates with the same probability and follows the same strategy of quality choices conditional on participating. We will denote a pair of participation probability and CDF that constitute a symmetric mixed strategy by $(p, G(q))$.

**Definition 2.2.** *A symmetric mixed strategy equilibrium $(p, G(q))$ is a probability p and a distribution G over qualities q such that when every agent enters with probability p, and chooses a quality drawn from the CDF $G(q)$ conditional on entering, no agent can increase her expected utility by deviating from this strategy, i.e., by changing either the probability with which she participates or the distribution G from which she draws a quality.*

# 3 Equilibrium Analysis

We begin by analyzing the equilibria of the mechanism $\mathcal{G}(a_1, a_2, \ldots, a_n)$, which we will then use to compare outcomes in different mechanisms. We first prove the following simple lemma, which eliminates the possibility of 'pure strategy' equilibria in which all participants choose the same quality.

**Lemma 3.1.** *There exists no symmetric equilibrium in the game $\mathcal{G}(a_1, a_2, \ldots, a_n)$ where all participants choose the same quality $q$ conditional on entering when the cost $c(q)$ is continuous.*

*Proof.* Suppose there is a symmetric equilibrium in which all participants choose the same quality $q$ conditional on entering. If $k$ agents enter (where $k$ can be a random variable if participants randomized over the choice of entry), the expected payoff to each agent that enters (where the expectation is over random tiebreaking) is $\sum_{i=1}^{k} a_i/k$. So the expected payoff to an agent who enters with quality $q$ is the expectation over all possible values of $k$,

$$E[U(q, q_{-i})] = \sum_{k=1}^{n} \Pr(k) \sum_{i=1}^{k} \frac{a_i}{k} - c(q).$$

But note that entering with quality $q + \epsilon$ is a profitable deviation if all other agents who enter choose quality $q$: choosing $q + \epsilon$ gives this agent a reward of $a_1$ for all values of $k$. Since $a_i > a_{i+1}$ for some $i$, $a_1 > \sum_{i=1}^{n} \frac{a_i}{n}$ (and $\Pr(n) > 0$ in a symmetric equilibrium in which all participants enter with some positive probability). Since $c(q)$ is continuous, there exists a choice of $\epsilon$ such that

$$E[U(q + \epsilon, q_{-i})] > E[U(q, q_{-i})]$$

which constitutes a profitable deviation, contradicting the assumption that there is a symmetric equilibrium in which all agents choose the same quality. $\square$

Since there can exist no symmetric pure strategy equilibria in which all agents choose a single quality $q^*$ conditional on entering, we will investigate symmetric mixed strategy equilibria where all agents randomize over their choice of quality (conditional on entering) using the same distribution. First, of course, we need to establish the existence of such symmetric mixed strategy equilibria—we will do this by explicitly constructing such an equilibrium.

The next theorem establishes some properties that any symmetric mixed strategy equilibrium to $\mathcal{G}(a_1, a_2, \ldots, a_n)$, if one exists, must possess. We will use these properties to prove the existence of a symmetric mixed strategy equilibrium by constructing one in Theorem 3.2.

**Theorem 3.1.** *Let $(p, G(q))$ be any symmetric mixed strategy equilibrium to $\mathcal{G}(a_1, a_2, \ldots, a_n)$. If the agents' cost $c(q)$ is continuous and strictly increasing in $q$, then $G(q)$ is continuous, i.e., contains no mass points, and has support on an interval with left endpoint 0.*

*Proof.* Let $U(q) = \sum_{i=1}^{n} a_i Pr(i|q)$, where $Pr(i|q)$ is the probability of being ranked $i$th when choosing quality $q$, given that the remaining agents participate with probability $p$ and draw qualities according to the distribution $G(q)$ conditional on participating, and ties are broken at random. $U(q)$ is the benefit to this agent from entering with quality $q$ when other agents play $(p, G(q))$. The payoff to an agent who enters with quality $q$, when other agents play according to a mixed strategy $(p, G(q))$ in $\mathcal{G}(a_1, a_2, \ldots, a_n)$ is

$$\pi(q) = U(q) - c(q) = \sum_{i}^{n} a_i Pr(i|q) \; - \; c(q).$$

1. $G(q)$ has no mass points: We first show that $G(q)$ is continuous on its support, *i.e.*, it has no mass points. Suppose not; let $q_0$ be a mass point. Then, since ties are broken randomly, note that

$$\lim_{\epsilon \to 0} U(q_0 + \epsilon) > U(q_0),$$

since there is a positive probability of a tie at $q = q_0$, which can be eliminated by choosing a slightly higher quality. This implies that there is an $\epsilon$ such that the payoff from choosing $q_0 + \epsilon$ is strictly greater than that at $q_0$, since the cost function $c$ is continuous in $q$:

$$\begin{aligned} \lim_{\epsilon \to 0} \pi(q_0 + \epsilon) &= \lim_{\epsilon \to 0} (U(q_0 + \epsilon) - c(q_0 + \epsilon)) \\ &> U(q_0) - c(q_0). \end{aligned}$$

But this means $q_0$ cannot belong to the support of an equilibrium distribution $G$, a contradiction. So $G(q)$ is continuous on its support.

2. $q_{\min} = 0$: Let $q_{\min}$ denote the infimum of the qualities in the support of $G$. Since $G$ contains no mass points as shown above, $G(q_{\min}) = 0$. Suppose, for a contradiction, that $q_{\min} > 0$. But then, an agent can profitably deviate by choosing $q_{\min} - \epsilon$ instead of $q_{\min}$: since $G(q_{\min}) = 0$, an agent choosing $q_{\min}$ will be ranked lowest among all agents who enter anyway, *i.e.*, $U(q_{\min} - \epsilon) = U(q_{\min})$. But $c(q_{\min}) > c(q_{\min} - \epsilon)$ since $c(q)$ is strictly increasing, so $\pi(q_{\min} - \epsilon) > \pi(q_{\min})$, yielding a profitable deviation, contradicting $q_{\min}$ belonging to the support of an equilibrium distribution. This argument holds for any quality strictly greater than the lowest quality, which is 0. Therefore, any equilibrium distribution $G$ must have $q_{\min} = 0$.

3. Interval support: Finally, we argue that the support of $G(q)$ must be an interval (*i.e.*, the support contains no 'holes'), or equivalently, $G(q)$ is strictly increasing between 0 and $\bar{q}$, where $G(\bar{q}) = 1$. Suppose not; then there must exist some $q_1 < q_2$ such that $G(q_1) = G(q_2)$ (recall that we have already ruled out mass points, so specifically, there can be no mass point at $q_2$). But then $U(q_1) = U(q_2)$, since the quality $q$ affects the probability of being ranked in any particular position only via $G(q)$ (see (1)). So

$$\pi(q_2) = U(q_2) - c(q_2) < U(q_1) - c(q_1) = \pi(q_1)$$

since $c(q_1) < c(q_2)$, a contradiction to $q_2$ belonging to the support in equilibrium. Therefore, there can be no such holes in the support of $G$, i.e., the support is an interval.

$\square$

The probability $Pr(i|q)$ that an agent choosing quality $q$ has the $i$th highest quality when the remaining $n-1$ agents play according to $(p, G(q))$ when $G(q)$ is *continuous*, is the probability that $i-1$ other agents participate (each with probability $p$) and choose quality greater than $q$ (each with probability $1-G(q)$), and the remaining $n-i$ agents either do not participate or choose quality less than $q$, i.e.,

$$Pr(i|q) = \binom{n-1}{i-1}(p(1-G(q)))^{i-1}(1-p(1-G(q)))^{n-i}.$$

Note that this expression is valid only because $G$ has no mass points, since if there is a mass point at $q$ there is a positive probability of more than one agent using the same quality, which leads to ties that are broken randomly.

Then, the benefit $U = \sum_{i=1}^{n} a_i Pr(i|q)$ under a *continuous* CDF $G$ is

$$U = \sum_{i=0}^{n-1} a_{i+1}\binom{n-1}{i}(p(1-G(q)))^{i}(1-p(1-G(q)))^{n-1-i}. \tag{1}$$

Before proceeding with the construction of an equilibrium, we evaluate a derivative which will be used repeatedly in our proofs in this section. Setting

$$x(q) = p(1-G(q)),$$

write

$$U(x) = \sum_{i=0}^{n-1} a_{i+1}\binom{n-1}{i}x^{i}(1-x)^{n-1-i}. \tag{2}$$

Then,

$$\frac{dU}{dx} = \sum_{i=0}^{n-1} a_{i+1}\binom{n-1}{i}ix^{i-1}(1-x)^{n-1-i} - \sum_{i=0}^{n-1} a_{i+1}\binom{n-1}{i}(n-1-i)x^{i}(1-x)^{n-2-i}$$

$$= (n-1)\sum_{i=0}^{n-2} a_{i+2}\binom{n-2}{i}x^{i}(1-x)^{n-2-i} - (n-1)\sum_{i=0}^{n-2} a_{i+1}\binom{n-2}{i}x^{i}(1-x)^{n-2-i},$$

so that

$$\frac{dU}{dx} = (n-1)\sum_{i=0}^{n-2}(a_{i+2} - a_{i+1})\binom{n-2}{i}x^{i}(1-x)^{n-2-i}. \tag{3}$$

For a monotone mechanism, *i.e.*, with $a_i \geq a_{i+1}$ and at least one strict inequality, note that $\frac{dU}{dx} > 0$ for $x \in (0,1)$.

Now we will use Theorem 3.1 to construct, and therefore demonstrate the existence of a symmetric mixed strategy equilibrium $(p, G(q))$ in $\mathcal{G}(a_1, a_2, \ldots, a_n)$.

**Theorem 3.2** (Equilibrium Construction). *There exists a symmetric mixed strategy equilibrium $(p, G(q))$ to $\mathcal{G}(a_1, a_2, \ldots, a_n)$ when $a_i \geq a_{i+1}$ for all $i$; this equilibrium is unique up to inclusion of the endpoints of the support.*

*Proof.* We will construct a candidate pair $(p, G(q))$ for which no agent can benefit by changing $p$ and no agent will want to deviate from $G(q)$; to finish the proof we verify that $p$ and $G(q)$ are indeed a valid probability and CDF, respectively.

Before constructing the equilibrium, we note that if $a_1 \leq c(0)$, *i.e.*, the maximum possible reward is less than the cost of producing the lowest possible quality, no agent can derive nonnegative utility from participating in $\mathcal{G}(a_1, a_2, \ldots, a_n)$ irrespective of the actions of other agents. In this case, the only equilibrium is that no agents participate in $\mathcal{G}(a_1, a_2, \ldots, a_n)$ (*i.e.*, $p = 0$; the choice of $G(q)$ is meaningless), which is not very interesting. In what follows, therefore, we will assume that $a_1 > c(0)$.

First, from Theorem 3.1, we know that a mixed strategy equilibrium $(p, G(q))$, if one exists, has support on an interval $[0, \bar{q}]$ with $G(0) = 0$ and $G(\bar{q}) = 1$. Also, since $G(q)$ is continuous, the payoff at quality $q \in [0, \bar{q}]$ is

$$\pi(q) = U(p(1 - G(q))) - c(q),$$

where $U(x)$ is the function defined in (2).

Using the fact that 0 belongs to the support and $G(0) = 0$ (no mass points), we can write the payoff at 0 as

$$\begin{aligned}\pi(0) &= U(p) - c(0) \\ &= \sum_{i=0}^{n-1} a_{i+1}\binom{n-1}{i}(1-p)^{n-1-i}p^i - c(0).\end{aligned}$$

If $p = 1$, $U(p) = U(1) = a_n$[2]. Therefore, if $p = 1$, $\pi(0) = a_n - c(0)$. But for $p$ to be an equilibrium probability of participation, we must have $\pi(0) \geq 0$. Therefore, $p$ can be 1 only if $a_n \geq c(0)$, *i.e.*, if $a_n < c(0)$ then $p < 1$ in equilibrium. Conversely, if $a_n \geq c(0)$, we must have $p = 1$ in equilibrium. At $p = 1$, $\pi(0) = a_n - c(0) \geq 0$. Since $U(p)$ is a strictly decreasing function of $p$ on $(0,1)$, $\pi(0) > 0$ for any $p < 1$. But then if $p < 1$, an agent has an incentive to deviate and increase the probability of participation since payoffs are strictly positive, contradicting the fact that $p$ is an equilibrium participation probability. Therefore, $p = 1$ if and only if $a_n \geq c(0)$.

---
[2]This corresponds to the fact that when all agents participate (p=1), an agent choosing 0 quality comes in last (recall that there is no mass point at 0) and gets benefit $a_n$.

For $(p, G(q))$ to be an equilibrium, we must have equal payoffs throughout the support, *i.e.*,
$$\pi(q) = K$$
for all $q \in [0, \bar{q}]$. Further, since $(p, G(q))$ is a free entry equilibrium, no agent must have an incentive to change her decision to participate. This means that if $p < 1$, we must have $K = 0$, *i.e.*, equilibrium payoffs must be zero unless $p = 1$. We now use this together with the previous argument to construct our equilibrium.

1. $a_n < c(0)$: If $a_n < c(0)$, then set $p$ to be the value that satisfies
$$\sum_{i=0}^{n-1} a_{i+1} \binom{n-1}{i} (1-p)^{n-1-i} p^i = c(0). \qquad (4)$$

    Note that the left-hand side is a continuous, strictly decreasing function of $p$ on $(0, 1)$, taking value $a_1$ at $p = 0$ and $a_n$ at $p = 1$. Therefore there is a unique solution in $(0, 1)$ to this equation when $c(0)$ satisfies $a_n < c(0) < a_1$, *i.e.*, there is a unique solution $p$ which is a valid probability.

    The distribution $G(q)$ is the solution to
$$\sum_{i=0}^{n-1} a_{i+1} \binom{n-1}{i} (1-p(1-G(q)))^{n-1-i} (p(1-G(q)))^i = c(q).$$

    for each $q$ in $[0, \bar{q}]$, where $\bar{q}$ is the unique solution (since $c(q)$ is strictly increasing) to
$$c(\bar{q}) = a_1.$$

    Note that the value of $\bar{q}$ is that which solves $G(\bar{q}) = 1$ in the equation above.

2. $a_n \geq c(0)$: If $a_n \geq c(0)$, set $p = 1$. The distribution $G(q)$ has support on the interval $[0, \bar{q}]$, where
$$c(\bar{q}) = a_1 - a_n + c(0),$$

    and $G(q)$ is given by the solution to
$$\sum_{i=0}^{n-1} a_{i+1} \binom{n-1}{i} (1-p(1-G(q)))^{n-1-i} (p(1-G(q)))^i = c(q) + a_n - c(0)$$

    for each $q$ in $[0, \bar{q}]$. Again, note that $\bar{q}$ is obtained by setting $G(\bar{q}) = 1$ in the equation above.

To verify that our construction is indeed an equilibrium, note that no agent has an incentive to deviate and choose a different $p$: when $p < 1$, $\pi(q) = 0$ so there is no benefit from increasing or decreasing $p$, and $\pi(q) \geq 0$ for $p = 1$ so no agent wants to decrease participation in this case. Also, no agent wants to deviate from $G(q)$ conditional on participating: first, $\pi(q)$ is equal for all $q \in [0, \bar{q}]$, so an agent might only want to deviate by choosing quality greater than $\bar{q}$. But note that in both cases ($p < 1$ and $p = 1$), $\pi(q) = U(\bar{q}) - c(q) < \pi(\bar{q})$ for any $q > \bar{q}$, since an agent choosing $\bar{q}$ is guaranteed to win the maximum possible reward anyway (recall that $G$ has no mass points, specifically at $\bar{q}$). So no agent wants to deviate from $(p, G(q))$.

We have already verified that the value of $p$ lies between 0 and 1, i.e., it is a valid probability. The last thing we need to verify is that the distribution $G(q)$ computed in both cases is indeed a CDF (note that the claimed properties of $G$, namely continuity with support on $[0, \bar{q}]$ follow directly from the continuity of $U(x)$ in $x$ and $c(q)$, and by construction). To show this, we need to show that that $G$ is increasing on $(0, \bar{q})$, i.e.,

$$\frac{\partial G}{\partial q} = \left(\frac{\partial U}{\partial q}\right) / \left(\frac{\partial U}{\partial G}\right)$$

is nonnegative on $(0, \bar{q})$. Now, observe that in either case ($a_n \geq c(0)$ or $a_n < c(0)$), $G(q)$ can be written as the solution to

$$U(p(1 - G(q))) = c(q) + \max\{a_n - c(0), 0\},$$

where $p$ is determined appropriately. Therefore, $U(q) = c(q) + \max\{a_n - c(0), 0\}$ is a strictly increasing function of $q$, i.e., $\frac{\partial U}{\partial q} > 0$. Also, with $x = p(1 - G)$,

$$\frac{\partial U}{\partial G} = \frac{\partial U}{\partial x} \cdot \frac{\partial x}{\partial G} = -p\frac{\partial U}{\partial x}.$$

Using the expression in (3), and the fact that $a_i \geq a_{i+1}$ with strict inequality for some $i$, $\frac{\partial U}{\partial G} > 0$ on $(0, \bar{q})$. So we have

$$\frac{\partial G}{\partial q} = \left(\frac{\partial U}{\partial q}\right) / \left(-p\frac{\partial U}{\partial x}\right) > 0$$

on $(0, \bar{q})$ (recall that $a_1 > c(0)$ by assumption, so $p > 0$). By construction, $G(0) = 0$ and $G(\bar{q}) = 1$, so $G(q)$ is increasing and lies in $[0, 1]$ for $q \in [0, \bar{q}]$. So $G(q)$ is a valid CDF. $\square$

The following two facts about the equilibrium are immediate from the proof above.

**Corollary 3.1.** *For any rewards $(a_1, \ldots, a_n)$ such that $a_i \geq a_{i+1}$ and at least one inequality is strict,*

1. *The equilibrium participation probability $p$ in $\mathcal{G}(a_1, a_2, \ldots, a_n)$ is 1 if and only if $a_n \geq c(0)$.*

2. *The maximum quality $\bar{q}$ in the support of $G$ is given by $c(\bar{q}) = a_1 - \max\{a_n - c(0), 0\}$.*

# 4   Increasing Attention Rewards

We begin with investigating the design of incentives in the context of attention rewards. Such attention rewards arise, for example, in sites that are based on user-generated content (UGC) such as Q&A forums like Quora or StackOverflow, or Amazon reviews. In these settings, there is some available amount of attention reward for the top 'spot' or answer (derived from all the viewers who read the contribution displayed first), a smaller amount for the second spot (corresponding to the viewers that continue on to the second), and so on, *i.e.*, some maximum possible rewards $A_1, A_2, \ldots, A_n$ that can be obtained by always showing all available contributions for each position $1, \ldots, n$.

Attention rewards have an unusual constraint when contrasted with monetary or virtual points rewards of the kind we discuss in §5: the total available reward $\sum_{i=1}^{n} A_i$ cannot be arbitrarily redistributed amongst agents since, to a first approximation, attention to the second spot comes from a subset of viewers providing attention to the first. Thus, while it is possible to freely increase or decrease each of the rewards $a_i$ between 0 and $A_i$ (subject, of course, to the monotonicity constraint, *i.e.*, $a_i \geq a_{i+1}$), it is not easy to take away reward from $a_2$ and redistribute it to $a_1$.[3]

Now, a site featuring UGC could suppress some of the UGC, e.g. by only showing the top-ranked content, or reducing the prominence of lesser ranked content, *i.e.*, the site could choose $a_i < A_i$ by not always (or never) displaying the $i$th ranked contribution. Does this strategy improve the quality of the best contribution supplied? On the one hand, equilibrium qualities should rise, because the payoff to poor quality falls. However, the payoff to supplying any content also falls, so participation falls as well. How do these two effects interact?

What if we were interested in a different metric of performance, and not just in the best contribution— for example, do the qualities of the average contribution, or the quality of the second best or third best contribution behave the same way as the quality of the best contribution as a function of $a_i$, or do they behave differently? Intuitively, it seems plausible that the solution for maximizing the quality of the best contribution may differ from what maximizes the quality of an average contribution, since lower rewards for non-winning contributions should increase the incentive to be best, but higher rewards for non-winning contributions may increase the average quality.

The following theorem says that the *entire* distribution of equilibrium qualities (accounting for the fact that agents participate probabilistically), and therefore every increasing statistic, improves when the rewards for achieving any of the the first through last-but-one ranks increases (this uniform improvement is in contrast to the case with redistribution, as we will see in §5). Therefore, it is optimal to increase each of the $a_1, a_2, \ldots, a_{n-1}$ to the maximum extent possible. However, the situation is somewhat more subtle for $a_n$, the subsidy to the contributor with the lowest possible rank: if the current value of $a_n < c(0)$ then increasing $a_n$ improves quality, but if $a_n$ is fairly large already, i.e., $a_n > c(0)$ then a decrease in $a_n$ improves quality.

---

[3] We note that randomizing between displaying $q_1$ and $q_2$ in the first and second spot can achieve the opposite redistribution, namely increase $a_2$ at the expense of $a_1$, but we do not consider this here since it adversely affects the user experience. The analysis in §5 addresses this issue.

**Lemma 4.1.** *The derivative of the probability with which an agent chooses quality greater than $q$ in equilibrium with respect to $a_i$, $\frac{d(p(1-G))}{da_i}$, is positive for $i = 1, \ldots, n-1$ for all $q \in (0, \bar{q})$. The derivative with respect to $a_n$, $\frac{d(p(1-G))}{da_n}$ is positive when $a_n < c(0)$ but negative for $a_n > c(0)$ for all $q \in (0, \bar{q})$.*

*Proof.* We have from the equilibrium construction that

$$H(p(1-G), \mathbf{a}) \equiv U(p(1-G(q)) - c(q) - \max\{a_n - c(0), 0\} = 0, \qquad (5)$$

where $U(x)$ is the benefit function defined in (1).

Differentiating (5) gives us

$$\frac{d(p(1-G))}{da_i} = -\frac{\frac{\partial H}{\partial a_i}}{\frac{\partial H}{\partial p(1-G)}}.$$

We use the derivative $\frac{\partial U}{\partial x}$ calculated in (3) for the denominator, with $x = p(1-G)$:

$$\begin{aligned}\frac{\partial H}{\partial x} &= (n-1)\sum_{i=0}^{n-2}(a_{i+2} - a_{i+1})\binom{n-2}{i}x^i(1-x)^{n-2-i} \\ &< 0\end{aligned}$$

for $x \in (0, 1)$, since $a_i \leq a_{i+1}$ with at least one strict inequality. For $a_1, \ldots, a_{n-1}$,

$$\frac{\partial H}{\partial a_i} = \binom{n-1}{i-1}(p(1-G))^{i-1}(1-p(1-G))^{n-i} > 0$$

for $q \in (0, \bar{q})$. Therefore, $\frac{d(p(1-G))}{da_i} > 0$ everywhere, i.e., increasing the rewards for each of the first through $n-1$th positions always improves the equilibrium distribution.

For $a_n$, when $a_n < c(0)$ or equivalently $p < 1$,

$$\frac{\partial U}{\partial a_n} = (p(1-G))^{n-1} > 0,$$

on $(0, \bar{q})$, but when $a_n \geq c(0)$ (so that $p = 1$),

$$\frac{\partial U}{\partial a_n} = (p(1-G))^{n-1} - 1 < 0.$$

Therefore, $\frac{d(p(1-G))}{da_n} > 0$ for $a_n < c(0)$ but $\frac{d(p(1-G))}{da_n} < 0$ for $a_n \geq c(0)$. Thus, the reward for the last position behaves differently— increasing $a_n$ until it equals $c(0)$ improves equilibrium qualities, but when $a_n \geq c(0)$, increasing $a_n$ further make the equilibrium qualities worse. □

Recall also from Corollary 3.1 that $c(\bar{q}) = a_1 - \max\{a_n - c(0), 0\}$, so that the maximum quality in the support decreases linearly with $a_n$ when $a_n \geq c(0)$.

This immediately gives us the following result.

**Theorem 4.1.** *Suppose each of the rewards $a_i$ is constrained to lie below some maximum value $A_i$, $0 \leq a_i \leq A_i$, where $A_1 \geq \ldots \geq A_n$. Then, the choice of rewards $(a_1, \ldots, a_n)$ that optimizes the equilibrium distribution of qualities, and therefore the expected value of any increasing function of the contributed qualities, is*

$$\begin{aligned} a_i &= A_i, \quad i = 1, \ldots, n-1; \\ a_n &= \min(A_n, c(0)). \end{aligned}$$

## 5 Redistribution of Rewards

We now address the question of how to optimally redistribute reward amongst agents to improve equilibrium quality. This question arises in settings where there is some total available reward that can be distributed in any arbitrary way amongst agents, as in the case of crowdsourcing contests such as TopCoder, or even contests with virtual points, where points have value only relative to the total number of points in the system, so that effectively there is a fixed budget of available reward. We note that this setting is the one that has been studied widely in the contest design literature in economics, and in the growing literature on the design of crowdsourcing contests, unlike the setting in §4; the key difference, as discussed in the section on related work, is that our model allows for *endogenous* entry. Which value of $(a_1, \ldots, a_n)$ leads to the 'best' equilibrium outcome amongst all mechanisms $\mathcal{G}(a_1, a_2, \ldots, a_n)$ with the same expected payout?

What do we mean by 'best' outcome, *i.e.*, what is the objective to optimize? As we will see, unlike in the previous section with attention rewards, not all increasing statistics of the quality distribution need be optimized by the same allocation of rewards. We will focus largely on the expected quality of the best contribution, since this is the objective of interest in many settings like crowdsourcing contests with an arbitrarily redistributable total reward, and finally briefly address the expected total quality, which is potentially relevant in settings like Q&A forums such as Y! Answers.

We first write the budget constraint that says we are restricted to redistributing rewards, *i.e.*, the total expected payout to contestants must remain the same. Since entry is endogenous, the number of participants in equilibrium is a random variable when $p < 1$, so not all prizes $a_i$ are always paid out. The expected payment to the winners in equilibrium is

$$B = \sum_{j=1}^{n} \binom{n}{j} p^j (1-p)^{n-j} \sum_{k=1}^{j} a_k$$

since the payment when $j$ contributors enter, which happens with probability $\binom{n}{j}p^j(1-p)^{n-j}$ where $p$ is the equilibrium participation probability, is $\sum_{k=1}^{j} a_k$. Rearranging, we have

$$B = \sum_{k=1}^{n} a_k \sum_{j=k}^{n} \binom{n}{j} p^j(1-p)^{n-j}. \tag{6}$$

Note that when $p = 1$, $B = \sum_{i=1}^{n} a_i$.

Before deriving our results for the maximum quality, we state a couple of technical lemmas. The proof of the first proposition below is obtained easily by integrating by parts.

**Proposition 5.1.** *For any $k \leq n$, and $p \geq 0$,*

$$\sum_{j=k}^{n} \binom{n}{j} p^j(1-p)^{n-j} = n\binom{n-1}{k-1} \int_0^p x^{k-1}(1-x)^{n-k} dx.$$

We introduce some notation before our next lemma.

**Definition 5.1** ($\mathcal{B}_k(q), \mathcal{W}(k)$). *Consider $n$ agents playing according to the symmetric mixed strategy $(p, G(q))$. We define*

$$\mathcal{B}_k(q) = \binom{n-1}{k-1}(p(1-G(q)))^{k-1}(1-p(1-G(q)))^{n-k};$$

$\mathcal{B}_k(q)$ *is the probability that an agent entering and choosing quality $q$ is ranked at position $k$. We also define*

$$\mathcal{W}(k) = \int_0^{\bar{q}} \mathcal{B}_k(q) p G'(q) dq.$$

$\mathcal{W}(k)$ is the probability that a particular one of the $n$ agents who enter with probability $p$ and choose quality from the distribution $G(q)$ is ranked in the $k$th position.

The proof of the following proposition uses the identity in Proposition 5.1:

**Proposition 5.2.** *For any index $s \leq n$,*

$$(1-(1-p)^n) \cdot \binom{n-1}{s-1} p^{s-1}(1-p)^{1-s} \geq \sum_{j=s}^{n} \binom{n}{j} p^j(1-p)^{n-j}.$$

The following technical lemma uses Proposition 5.2 above, and is central to the proof of the main lemma.

**Lemma 5.1.** *Suppose $a_n < c(0)$, i.e., $p < 1$, and we vary the reward $a_s$ for some rank $s$ and change $a_1$ to keep the budget $B$ unchanged.*

$$\left. \frac{da_1}{da_s} \right|_B \leq -\frac{\mathcal{W}(s)}{\mathcal{W}(1)}.$$

*Proof.*
$$\left.\frac{da_1}{da_s}\right|_B = -\frac{\partial B}{\partial a_s} \Big/ \frac{\partial B}{\partial a_1}.$$

We first evaluate the quantity $\frac{\partial B}{\partial a_s}$ when $p < 1$:

$$\begin{aligned}
\frac{\partial B}{\partial a_s} &= \sum_{j=s}^{n}\binom{n}{i}p^i(1-p)^{n-j} + \sum_{k=1}^{n}a_k\sum_{j=k}^{n}\binom{n}{i}\left(jp^{j-1}(1-p)^{n-j} - (n-j)p^i(1-p)^{n-j-1}\right)\frac{dp}{da_s}\\
&= \sum_{j=s}^{n}\binom{n}{i}p^i(1-p)^{n-j} + n\sum_{k=1}^{n}\binom{n-1}{k-1}p^{k-1}(1-p)^{n-k}\frac{dp}{da_s},
\end{aligned}$$

which, using (4), gives us

$$\frac{\partial B}{\partial a_s} = \sum_{j=s}^{n}\binom{n}{i}p^i(1-p)^{n-j} + nc(0)\frac{dp}{da_s}. \tag{7}$$

Now, note that

$$\frac{dp}{da_s} = \frac{\binom{n-1}{s-1}p^{s-1}(1-p)^{n-s}}{(n-1)\sum_{i=0}^{n-2}(a_{i+2}-a_{i+1})\binom{n-2}{i}p^i(1-p)^{n-2-i}}.$$

Using both of these, together with the inequality in Proposition 5.2 and the definition of $\mathcal{W}(k)$, and rearranging, gives the result. □

We now state and prove the main lemma in this section, which will immediately give us the theorem on maximizing the expected quality of the best contribution.

**Lemma 5.2.** *Suppose the cost function $c$ is such that $c'(q)/c(q)$ is non-increasing in $q$. Then, for any monotone nonnegative contests $\mathcal{G}(a_1, a_2, \ldots, a_n)$, redistributing reward away from the winner to any lower rank $s > 1$ locally decreases the expected quality of the best contribution in equilibrium, i.e.,*

$$\left.\frac{dEq_{\max}}{da_s}\right|_{B,a_1} \leq 0.$$

*Proof.* The expected value of the highest quality contribution obtained in an equilibrium of $\mathcal{G}(a_1, a_2, \ldots, a_n)$, counting the utility from receiving no contributions as the same as from a zero quality contribution, is

$$Eq_{\max} = \int_0^{\bar{q}} 1 - (1 - p(1 - G(q)))^n \, dq.$$

We are interested in the effect of shifting reward from the winner to some lower rank $s$ on the expected highest quality contribution in equilibrium, *i.e.*, the effect of changing $a_s$ when $a_1$ is adjusted so as to preserve the expected payout on $Eq_{\max}$:

$$\begin{aligned}\left.\frac{dEq_{\max}}{da_s}\right|_{B,a_1} &= \left.\frac{d}{da_s}\int_0^{\bar{q}} 1-(1-p(1-G(q)))^n dq\right|_{B,a_1} \\ &= \int_0^{\bar{q}} np(1-p+pG(q)))^{n-1}\left\{\frac{dp(1-G(q))}{da_s}+\left.\frac{da_1}{da_s}\right|_B \frac{dp(1-G(q))}{da_1}\right\} dq.\end{aligned}$$

Recall the equilibrium condition from (5):

$$H(p(1-G),\mathbf{a}) = U(p(1-G(q))-c(q)-\max\{a_n-c(0),0\} = 0.$$

Differentiating, we have

$$\frac{\partial H}{\partial p(1-G)}pG'(q) = -c'(q). \qquad (8)$$

Therefore,

$$\frac{dp(1-G(q))}{da_k} = -\frac{\frac{\partial H}{\partial a_i}}{\frac{\partial H}{\partial p(1-G)}} = \frac{\mathcal{B}_k(q)pG'(q)}{c'(q)}, \qquad (9)$$

where $\mathcal{B}_k(q)$ is as in Definition 5.1.

**Case 1: $a_n < c(0)$, or $p < 1$.** Using the inequality bounding $\frac{da_1}{da_s}$ from Lemma 5.1, we have

$$\begin{aligned}\left.\frac{dEq_{\max}}{da_s}\right|_{B,a_1} &\leq \int_0^{\bar{q}} np(1-p+pG(q)))^{n-1}\left\{\mathcal{B}_s(q) - \frac{\mathcal{W}(s)}{\mathcal{W}(1)}\mathcal{B}_1(q)\right\}\frac{pG'(q)}{c'(q)}dq \\ &= n\mathcal{W}(s)\int_0^{\bar{q}}\frac{(1-p+pG(q)))^{n-1}}{c'(q)}\left\{\frac{\mathcal{B}_s(q)pG'(q)}{\mathcal{W}(s)} - \frac{\mathcal{B}_1(q)pG'(q)}{\mathcal{W}(1)}\right\}dq.\end{aligned}$$

Now, recall that each term multiplying $\frac{(1-p+pG(q)))^{n-1}}{c'(q)}$ in this difference is a density:

$$f_k(q) = \frac{\mathcal{B}_k(q)pG'(q)}{\mathcal{W}(k)} = \frac{\mathcal{B}_k(q)pG'(q)}{\int_0^{\bar{q}}\mathcal{B}_k(q)pG'(q)dq}.$$

Therefore, the term within the parentheses is the difference between two densities $f_s(q)$ and $f_1(q)$, with the property that the first density (corresponding to $s$) puts less weight on higher values of $q$ (formally, it is easy to verify that the two distributions $f_1$ and $f_s$ satisfy the MLRP property if $s > 1$, which implies first order stochastic dominance). Therefore, the right-hand side is the difference between the expected value of the function $\frac{(1-p+pG(q)))^{n-1}}{c'(q)}$ computed with respect to the densities $f_s(q)$ and $f_1(q)$. Therefore, $\frac{dEq_{\max}}{da_s}$ is non-positive if $\frac{(1-p+pG(q)))^{n-1}}{c'(q)}$ is an increasing function of $q$ or equivalently that $\frac{c'(q)}{(1-p+pG(q)))^{n-1}}$ is decreasing. Recall that this derivative being non-positive implies that increasing the reward $a_1$, while decreasing $a_s$ in such a way as to hold the budget constant, improves the expected highest quality.

We now show that a sufficient condition for this is that $\frac{c'(q)}{c(q)}$ is non-increasing in $q$. The equilibrium condition gives us:

$$\frac{1}{c(q)} \sum_{j=0}^{n-1} \binom{n-1}{j} (p(1-G(q)))^j (1-p(1-G(q)))^{n-j-1} a_{j+1} = 1.$$

Using this, we have

$$\frac{c'(q)}{(1-p+pG(q))^{n-1}} = \frac{c'(q)}{c(q)} \sum_{j=0}^{n-1} \binom{n-1}{j} \frac{p(1-G(q))}{(1-p(1-G(q)))}^j a_{j+1}.$$

Since $\left(\frac{p(1-G(q))}{(1-p(1-G(q)))}\right)^j$ is decreasing in $q$ for every $j$ and each $a_{j+1} \geq 0$ by assumption, we have that if $\frac{c'(q)}{c(q)}$ is decreasing in $q$, then $\frac{c'(q)}{(1-p+pG(q))^{n-1}}$ is decreasing in $q$ as well, completing the proof.

**Case 2: $a_n \geq c(0)$, or $p = 1$.** In this case, all agents always participate, and the budget constraint simplifies to $B = \sum_{i=1}^n a_i$. Therefore,

$$\left.\frac{da_1}{da_s}\right|_B = -1.$$

Also, since $p = 1$, we have

$$\left.\frac{dEq_{\max}}{da_s}\right|_{B,a_1} = -\frac{d}{da_s}\int_0^{\bar{q}} 1 - G(q)^n dq \bigg|_{B,a_1}$$
$$= \int_0^{\bar{q}} nG(q)^{n-1}\left\{\frac{dG(q)}{da_s} + \frac{da_1}{da_s}\frac{dG(q)}{da_1}\right\} dq$$
$$= \int_0^{\bar{q}} nG(q)^{n-1}\left\{\frac{dG(q)}{da_s} - \frac{dG(q)}{da_1}\right\} dq.$$

Using (8) with $p = 1$, and noting that $\frac{\partial H}{\partial a_k} = \mathcal{B}_k(q)$ for $k = 1, \ldots, n-1$, we have as before

$$\frac{dG(q)}{da_k} = \frac{\mathcal{B}_k(q)G'(q)}{c'(q)},$$

for $k = 1, \ldots, n-1$. We substitute this to obtain for any $s < n$:

$$\left.\frac{dEq_{\max}}{da_s}\right|_{B,a_1} = \int_0^{\bar{q}} \frac{G(q)^{n-1}}{c'(q)} (n\mathcal{B}_s(q)G'(q) - n\mathcal{B}_1(q)G'(q)) dq.$$

Now, $n\mathcal{B}_k(q)G'(q)$, which is equal to $n\binom{n-1}{k-1}(1-G(q))^{k-1}G(q))^{n-k}G'(q)$ is positive and integrates out to 1, so it is a density. Moreover, it puts more weight on higher $q$ for lower $k$. Therefore, if $\frac{G(q)^{n-1}}{c'(q)}$ is increasing, $\left.\frac{dEq_{\max}}{da_s}\right|_{B,a_1}$ will be negative since $s > 1$. As before, we substitute

$$a_n - c(0) + \sum_{j=0}^{n-1} \binom{n-1}{j}(1-G(q)))^j G(q))^{n-j-1} a_{j+1} = c(q)$$

to obtain

$$\frac{c'(q)}{G(q)^{n-1}} = \frac{c'(q)}{c(q)}\left(\frac{c(0)}{G(q)^{n-1}} + \sum_{j=0}^{n-2}\binom{n-1}{j}\frac{(1-G(q))^j}{G(q)^j}a_{j+1}\right).$$

Again, the term within parentheses is decreasing in $q$, so if $\frac{c'(q)}{c(q)}$ is decreasing, then $\frac{G(q)^{n-1}}{c'(q)}$ is increasing, and the derivative $\left.\frac{dEq_{\max}}{da_s}\right|_{B,a_1}$ is nonpositive.

Finally, note that when $s = n$, $\frac{dG(q)}{da_s} = \mathcal{B}_k(q) - 1 \leq 0$, so that derivative can be immediately seen to be negative. Together, we have the result for the case $a_n \geq c(0)$. □

This lemma immediately gives us the two main theorems. The first result states that *if* we are restricted to nonnegative rewards $a_i$, *i.e.*, charging for entry is not feasible, then a winner-take-all contest maximizes the expected quality of the best contribution in equilibrium amongst all possible monotone, nonnegative allocations of the total budget amongst participants. This result agrees with the results from the literature on contest design and crowdsourcing contests which do not model endogenous entry (eg [1, 3]).

**Theorem 5.1.** *Suppose the cost function $c$ is such that $c'(q)/c(q)$ is nonincreasing in $q$. Then, the expected quality of the best contribution obtained in equilibrium among all monotone nonnegative contests $\mathcal{G}(a_1, a_2, \ldots, a_n)$ that have the same total expected payout, is maximized by a winner-take-all contest, i.e., at $(a_1^*, 0, \ldots, 0)$.*

The second result states that free entry does not lead to the optimum level of quality for the best contribution, and in fact restricting entry by taxation can improve the maximum equilibrium quality.

**Theorem 5.2.** *Consider a winner-take-all contest with rewards $(A, 0, \ldots, 0)$, and suppose $c(0) > 0$. If the cost $c$ is such that $c'(q)/c(q)$ is decreasing in $q$, then* taxing *entry locally improves the expected quality of the best contribution in equilibrium.*

Theorem 5.2 shows that it is advantageous to tax participants and use the proceeds to subsidize the best quality result, in order to maximize the best quality result. In models where participation is exogenous, the desirability of a tax would not be surprising, because some additional profits can be extracted with no loss of participation. In contrast, with endogenous entry, a tax will drive down participation, which means we choose the maximum from a smaller number of random variables, potentially leading to a poorer outcome. It is therefore not surprising that the theorem on the optimality of a tax requires a condition on the cost function, although we note that this condition is satisfied by linear costs which are typically used in the crowdsourcing contest design literature [4, 1, 3], as well as exponential and other cost functions. What this condition accomplishes is to insure that the gain from improving the distribution of quality of participants will dominate the loss of participation from a small tax.

**Average or total quality.** What if we are interested in the average, or total, quality instead of the maximum quality? The average quality is simply the expected value of $q$ drawn according to the CDF $1 - p(1 - G(q))$, where as before, we count nonparticipation, or no contribution, as producing the same utility as a contribution with quality 0: $Eq_{\text{avg}} = \int_0^{\bar{q}} 1 - (1 - p(1 - G(q))) dq = \int_0^{\bar{q}} p(1 - G(q)) dq$. The total quality is $n$ times this average quality.

Here, unlike the case with attention rewards, we will see that the mechanism that is best for maximum quality need not be the best for average quality. We state the following two theorems.

**Theorem 5.3.** *Suppose $c'(q) = 1$, and let denote the expected total quality. Consider the winner-take-all contest $\mathcal{G}(a, 0, \ldots, 0)$. Then $\left.\frac{dEq_{\text{avg}}}{da_s}\right|_{B,a_1} \leq 0$ at $\mathcal{G}(a, 0, \ldots, 0)$.*

*Proof.* Since $a_n = 0$, and we have assumed that participation incurs a nonzero cost, i.e., $c(0) > 0$, we have $p < 1$. Then, the zero payoff equilibrium condition is
$$c(q) = a_1(1 - p(1 - G(q)))^{n-1}.$$

(Recall that we are considering winner-take-all contests.) Differentiating, and using the assumption of linear cost functions, we have

$$c'(q) = a_1(n-1)(1 - p(1 - G(q)))^{n-2} pG'(q). \tag{10}$$

We have

$$\left.\frac{dEq_{\text{avg}}}{da_s}\right|_{B,a_1} = \left.\frac{d}{da_s} \int_0^{\bar{q}} p(1 - G(q)) dq\right|_{B,a_1}$$
$$= \int_0^{\bar{q}} \left\{ \frac{dp(1 - G(q))}{da_s} + \left.\frac{da_1}{da_s}\right|_B \frac{dp(1 - G(q))}{da_1} \right\} dq.$$

Using (9) and setting $a_2 = \ldots = a_n = 0$, we have

$$\frac{dp(1 - G(q))}{da_s} = \frac{\binom{n-1}{s-1}(p(1 - G(q)))^{s-1}(1 - p(1 - G(q)))^{n-s}}{a_1(n-1)(1 - p(1 - G(q)))^{n-2}}.$$

Using the calculations in the proof of Lemma 5.1, and again using $a_2 = \ldots = a_n = 0$, we get

$$\left.\frac{dEq_{\text{avg}}}{da_s}\right|_{B,a_1} = \frac{1}{a_1(n-1)} \int_0^{\bar{q}} \binom{n-1}{s-1} (p(1-G(q)))^{s-1}(1-p(1-G(q)))^{2-s}$$

$$- \frac{\sum_{j=s}^n \binom{n}{j} p^j (1-p)^{n-j} + \frac{n}{n-1}\binom{n-1}{s-1} p^{s-1}(1-p)^{n+1-s}}{1-(1-p)^n + \frac{n(1-p)^n}{n-1}} (1-p(1-G(q))) dq$$

(using (10))

$$= \int_0^{\bar{q}} \left(\binom{n-1}{s-1}(p(1-G(q)))^{s-1}(1-p(1-G(q)))^{n-s} pG'(q)\right.$$

$$\left.- \frac{\sum_{j=s}^n \binom{n}{j} p^j (1-p)^{n-j} + \frac{n}{n-1}\binom{n-1}{s-1} p^{s-1}(1-p)^{n+1-s}}{1-(1-p)^n + \frac{n(1-p)^n}{n-1}} (1-p(1-G(q)))^{n-1} pG'(q)\right) dq$$

$$= \int_0^p \left(\binom{n-1}{s-1} x^{s-1}(1-x)^{n-s}\right.$$

$$\left.- \frac{\sum_{j=s}^n \binom{n}{j} p^j (1-p)^{n-j} + \frac{n}{n-1}\binom{n-1}{s-1} p^{s-1}(1-p)^{n+1-s}}{1-(1-p)^n + \frac{n(1-p)^n}{n-1}} (1-x)^{n-1}\right) dx$$

$$= \frac{1}{n}\sum_{j=s}^n \binom{n}{j} p^j (1-p)^{n-j} - \frac{\sum_{j=s}^n \binom{n}{j} p^j (1-p)^{n-j} + \frac{n}{n-1}\binom{n-1}{s-1} p^{s-1}(1-p)^{n+1-s}}{1-(1-p)^n + \frac{n(1-p)^n}{n-1}} \cdot \frac{1-(1-p)^n}{n}$$

$$= \frac{\frac{n}{n-1}(1-p)^n}{n(1+\frac{(1-p)^n}{n-1})} \left(\sum_{j=s}^n \binom{n}{j} p^j (1-p)^{n-j} - (1-(1-p)^n) \cdot \binom{n-1}{s-1} p^{s-1}(1-p)^{1-s}\right)$$

$$\leq 0,$$

where the final inequality follows from applying Proposition 5.2. $\square$

This theorem says that for linear cost functions, the equilibrium expected total quality is increased by increasing the reward to the highest rank at the expense of any lower rank at the winner-take-all contest $(a, 0, \ldots, 0)$. Here, there is too much entry for the average or total quality objective as well, and *taxation*, or charging entrants a small fee that is rebated to the winner, locally improves total quality.

Next, we consider exponential cost functions, $c(q) = e^{kq}$ $(k > 0)$— here, whether the average quality improves with taxes or subsidies depends on the size of the available reward $B$.

**Theorem 5.4.** *Suppose $c'(q)/c(q) = k$, where $k > 0$ is independent of $q$. Consider the winner-take-all contest $\mathcal{G}(a, 0, \ldots, 0)$. Then $\left.\frac{dEq_{\text{avg}}}{da_s}\right|_{B,a_1} < 0$ for small enough $B$, while $\left.\frac{dEq_{\text{avg}}}{da_s}\right|_{B,a_1} > 0$ for large enough $B$.*

*Proof.* Using the zero profit equilibrium condition and (10), we have

$$k = \frac{c'(q)}{c(q)} = \frac{(n-1)pG'(q)}{(1-p(1-G(q)))}.$$

We have

$$\left.\frac{dEq_{\text{avg}}}{da_s}\right|_{B,a_1} = \left.\frac{d}{da_s}\int_0^{\bar{q}} p(1-G(q))dq\right|_{B,a_1}$$

$$= \frac{1}{a_1(n-1)}\int_0^{\bar{q}} \frac{(n-1)pG'(q)}{(1-p(1-G(q)))}(\binom{n-1}{s-1}(p(1-G(q)))^{s-1}(1-p(1-G(q)))^{2-s}$$

$$- \frac{\sum_{j=s}^n \binom{n}{j}p^j(1-p)^{n-j} + \frac{n}{n-1}\binom{n-1}{s-1}p^{s-1}(1-p)^{n+1-s}}{1-(1-p)^n + \frac{n(1-p)^n}{n-1}}(1-p(1-G(q))))dq$$

$$= \frac{1}{a_1}\int_0^{\bar{q}}(\binom{n-1}{s-1}(p(1-G(q)))^{s-1}(1-p(1-G(q)))^{1-s}pG'(q)$$

$$- \frac{\sum_{j=s}^n \binom{n}{j}p^j(1-p)^{n-j} + \frac{n}{n-1}\binom{n-1}{s-1}p^{s-1}(1-p)^{n+1-s}}{1-(1-p)^n + \frac{n(1-p)^n}{n-1}}pG'(q))dq$$

$$= \frac{1}{a_1}\int_0^p \left(\binom{n-1}{s-1}x^{s-1}(1-x)^{1-s} - \frac{\sum_{j=s}^n \binom{n}{j}p^j(1-p)^{n-j} + \frac{n}{n-1}\binom{n-1}{s-1}p^{s-1}(1-p)^{n+1-s}}{1-(1-p)^n + \frac{n(1-p)^n}{n-1}}\right)dx$$

$$= \frac{1}{a_1}\left(\int_0^p \binom{n-1}{s-1}x^{s-1}(1-x)^{1-s}dx - p\frac{\sum_{j=s}^n \binom{n}{j}p^j(1-p)^{n-j} + \frac{n}{n-1}\binom{n-1}{s-1}p^{s-1}(1-p)^{n+1-s}}{1+\frac{(1-p)^n}{n-1}}\right).$$

Now let

$$f(p) = \int_0^p \binom{n-1}{s-1}x^{s-1}(1-x)^{1-s}dx - p\frac{\sum_{j=s}^n \binom{n}{j}p^j(1-p)^{n-j} + \frac{n}{n-1}\binom{n-1}{s-1}p^{s-1}(1-p)^{n+1-s}}{1+\frac{(1-p)^n}{n-1}},$$

and observe that $f(p)$ is positive for large $p$ and negative for small $p$ (recall $s > 1$). To see that $f(p)$ is negative for small $p$, first note that $f(0) = 0$. Also,

$$\lim_{p\to 0}\frac{f'(p)}{p^{s-1}} = \binom{n-1}{s-1} - \frac{\frac{n}{n-1}\binom{n-1}{s-1}}{\frac{n}{n-1}} - \frac{\frac{n}{n-1}\binom{n-1}{s-1}(s-1)}{\frac{n}{n-1}}$$

$$= -\binom{n-1}{s-1}(s-1)$$

$$< 0.$$

Therefore $f(p)$ is negative for small $p$, and

$$\left.\frac{dEq_{\text{avg}}}{da_s}\right|_{B,a_1} < 0$$

for $p$ near zero, corresponding to a small $B$. For large $p$,

$$f(p) \geq \int_0^p \binom{n-1}{s-1} x^{s-1}(1-x)^{1-s} dx - 1 - \frac{n}{n-1}\binom{n-1}{s-1} p^{s-1}(1-p)^{n+1-s}$$
$$\to \infty,$$

since for $s \geq 2$ (recall $x < 1$), $\int_0^p x^{s-1}(1-x)^{1-s} dx \geq \int_{1/2}^p (\frac{x}{1-x}) dx \to -\ln(1-p) \to \infty$ as $p \to 1$. Therefore,

$$\left.\frac{dEq_{\text{avg}}}{da_s}\right|_{B,a_1} > 0$$

when $p$ is large, corresponding to large $B$, for exponential costs. □

This theorem says that for exponential costs, the effect on the expected average quality of increasing $a_1$ while decreasing $a_s$ to maintain the budget for any $s > 1$, *depends* on the value of $B$: when $B$ is small, taxing entry improves average quality, but when $B$ is large, the average quality is increased by subsidizing entry. Recall that our results on the expected *maximum* quality do apply to exponential costs, and suggest that taxing entry is optimal for maximizing the expected quality of the best contribution. Thus, when the available reward $B$ is large, the mechanisms to maximize the quality of the best and average contributions need *not* be the same — taxing entry improves the best contribution's quality, whereas *subsidizing* entry is what improves the total quality of contributions produced over a winner-take-all contest for exponential cost functions and large $B$.

# 6  Acknowledgments.

We are grateful to Matt Jackson for helpful remarks, and in particular suggesting the idea of taxing entry to improve quality.